\documentclass[conference]{IEEEtran}
\IEEEoverridecommandlockouts

\usepackage{cite}
\usepackage{amsmath,amssymb,amsfonts}
\usepackage{algorithmic}
\usepackage{graphicx}
\usepackage{textcomp}
\usepackage{xcolor}
\usepackage{verbatim}
\usepackage{array}
\usepackage{url}
\usepackage{subfigure}
\usepackage{comment}
\usepackage{esdiff}
\usepackage{multirow}
\usepackage{multicol}
\usepackage[symbol]{footmisc}
\def\BibTeX{{\rm B\kern-.05em{\sc i\kern-.025em b}\kern-.08em
    T\kern-.1667em\lower.7ex\hbox{E}\kern-.125emX}}
    
\begin{document}

\title{Modeling and Characterization of Metastability in Single Flux Quantum (SFQ) Synchronizers}

\author{\IEEEauthorblockN{Gourav Datta, Peter A.~Beerel}
\IEEEauthorblockA{\textit{Ming Hsieh Department of Electrical and Computer Engineering} \\
\textit{University of Southern California}\\
Los Angeles, California 90089, USA \\
\{gdatta, pabeerel\}@usc.edu}
}

\maketitle

\begin{abstract}
Despite the promises of low-power and high-frequency of single-flux quantum (SFQ) technology, scaling these circuits remains a serious challenge that motivates the support of multiple SFQ clock domains. Towards this end, this paper analyzes the impact of setup time violations and metastability in SFQ circuits comparing the derived analytical models to their CMOS counterparts. 
It then extends this model to estimate the Mean Time Between Failure (MTBF) of flip-flop-based synchronizers and curve fits this model to simulations in the state-of-the-art SFQ5ee process. Interestingly, we find a two-flop SFQ synchronizer has an estimated MTBF of ${\sim}10^{6}$ years. 
\end{abstract}

\section{Introduction}\label{intro}

With CMOS technology facing increased challenges due to the limits of physical scaling \cite{end_moore}, superconductive digital electronics (SDE), especially single flux quantum (SFQ) \cite{likharev1991}, has appeared as a promising beyond-CMOS device technology supporting frequencies up to $370$ $GHz$ \cite{high_freq_sfq} and yielding switching energy per bit of $~10^{-19}J$ at $T=4.2K$ (liquid helium temperature) \cite{switch_energy1, switch_energy2}. Recently, several variants of SFQ technologies with even higher energy efficiency have been demonstrated \cite{mukhanov2011, mukhanov2, low_power_sfq, low_power_sfq1, low_power_sfq2}. Still, the promise of three orders of magnitude lower in power (in the case of non-resistive bias networks \cite{mukhanov2011}) at an order of magnitude higher frequency \cite{likharev1991}, has not yet been attained, primarily due to i) high process variations \cite{bunyk50, parameters1995}, ii) the lack of a compact and reliable memory element and controllable switch element, and iii) the lack of design automation methodologies and techniques that enable the design of large-scale SFQ circuits.

In particular, the 
ultra-high clock frequencies associated with SFQ makes low-skew clock distribution extremely challenging \cite{timingSpringer}. As a result, a $1 THz$ device was forced to function at a disastrous $20$ $GHz$ frequency \cite{bunyk50}. One approach to address this clocking challenge is to decompose the SFQ design into multiple blocks that are independently clocked, i.e., into multiple clock domains, similar to how large CMOS designs are managed. Since these clock domains have no phase relationship, no static timing constraints can be created for data transfer between them. As a result, the timing constraints between the flip-flops (FFs) at the boundary of these domains may be violated and the sampling FF in the receiver domain can exhibit metastability \cite{Metastability}.

To reduce the chance of metastable events propagating through a 
design, designers often use a sequence of back-to-back FFs, called a \textit{synchronizer} \cite{metastability_fpga, kinniment_tutorial}, whenever data is transferred between unrelated clock domains. Should the output of the first synchronization FF become metastable, it still needs to propagate through the rest of the sequence before its value is used by the rest of the design. The extra amount of time provided by the additional synchronization FFs increases the probability that the metastable value will resolve, and lowers the possibility that the design will fail \cite{metastable_performance}. The cost of the synchronizer is that it increases design latency.

This paper analyzes metastability in SFQ circuits and then quantifies it in the form of the Mean Time Between Failures (MTBF) of SFQ synchronizers that consist of a sequence of back-to-back FFs. To the best of our knowledge, we are the first to propose an analytical model of metastability in SFQ. We extract different parameters of this model using circuit simulations to compute MTBF in the current-state-of-the-art SFQ process, 
SFQ5ee \cite{sfq5ee}. We then discuss how 
multi-FF synchronizers can improve this MTBF. 

The remainder of the paper is organized as follows. Section \ref{sec:background} provides related background on SFQ, including a description of a SFQ FF used to store 
SFQ pulses and metastability in SFQ.
Section \ref{sec:Metastability_modeling}  derives an analytical model for metastability in SFQ logic circuits. Section \ref{sec:Metastability_simulate} performs JSIM \cite{jsim} simulations to generate the model parameters and compute MTBF of different FF-based synchronizers. Finally some conclusions are given in Section \ref{sec:conc}. 

\begin{figure*}[!t]
\centering {
\includegraphics[width=18.6cm]{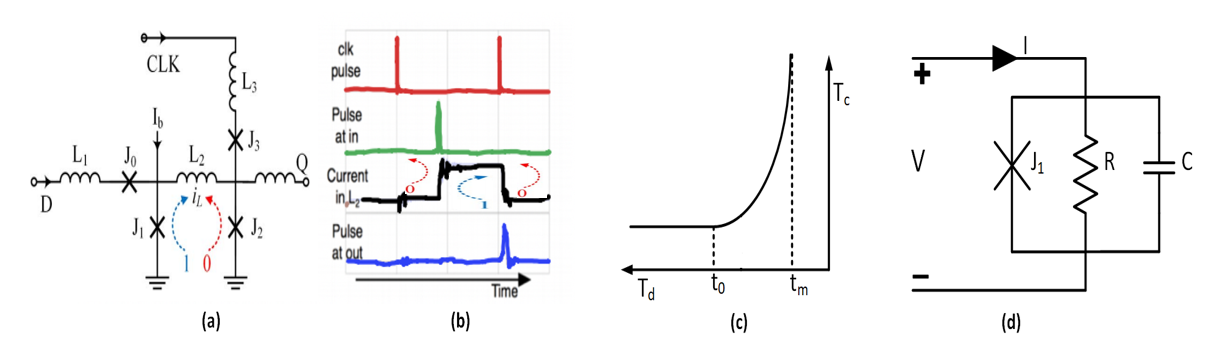}
 \vspace{-0.7cm}
\caption{(a) Schematic of a DFF (b) Simulation result of a DFF in SFQ (c) Illustration of clock-to-Q delay $T_c$ as a function of the 
arrival time of the data pulse $T_d$ relative to the clock. 
(d) RCSJ model of $J_1$}
\label{fig:dff}
}
 \vspace{-.4cm}
\end{figure*}

\section{Background}\label{sec:background}

\subsection{SFQ}

Unlike in CMOS, in SFQ technology, binary information is represented by very short (picosecond) voltage pulses $V(t)$ of quantized area, corresponding to transition of a single flux quantum, $\phi_{0}=\int V(t)dt=\frac{h}{2e}=2.03$ $mV.ps$.
These SFQ pulses can be quite naturally generated, reproduced, amplified, memorized, and processed by elementary cells comprising overdamped Josephson junctions (JJs) \cite{likharev1991}. 
In particular, the DC superconducting quantum interference device (SQUID) is the fundamental structure that is used as a memory element to store SFQ pulses \cite{bunyk50} and, to explain its use, we illustrate a SFQ D flip-flop (DFF) along with representative simulation waveforms in Figs. \ref{fig:dff}(a) and \ref{fig:dff}(b).  The DFF has two stable states, 0 or 1, that are characterized by the direction of the quantized current $i_{L}$ (one fluxon) in the loop consisting of two JJs, labelled $J_1$ and $J_2$, separated by an inductor $L_2$. Depending upon the state of the DFF, the arrival of clock pulse causes either the $J_2$ to leap (if the state is 1) or $J_3$ to leap (if the state is 0). If $J_2$ leaps, an output pulse will be generated losing the fluxon stored in the loop and resetting it to 0 state. On the other hand, if $J_3$ leaps, no pulse will be generated. $J_2$ and $J_3$ together form the Josephson comparator which senses the input current $i_L$ to decide which JJ to leap. Previous work has analyzed the switching characteristics of these comparators \cite{herr_2006,filippov_1999}, but to the best of our knowledge, there is no prior work modeling their output delay.

\subsection{Metastability in SFQ (Increased Clock-to-Q Delay)}

Because of the quantized nature of SFQ, there is no notion of a metastable voltage and there can never be a runt pulse generated at the output of an SFQ DFF. Either the output generates a pulse with energy of one fluxon or it does not. However, when the input pulse violates the setup time of the DFF, the clock-to-Q delay of the DFF can increase in an unbounded manner (see Fig. \ref{fig:dff}(c)), similar to what is observed in CMOS and bipolar technologies \cite{fangzhou,gdatta_isec,macromodular}\footnote{Interestingly, when the output pulse is delayed by a clock, the DFF exhibits the nominal clock-to-Q delay.}. As a result, this additional delay in the current pipeline stage can bleed into the next stage and cause a setup failure there \cite{fangzhou}. 

In CMOS, the clock-to-Q delay of a DFF accounts for a small portion of the clock period due to the presence of 6+ levels of logic gates \cite{cmos_depth} in the combinational path. In contrast, in SFQ, each logic gate is clocked, i.e., SFQ is inherently gate-level pipelined. This means that the number of clock sinks is large and the clock-to-Q delay is a dominant factor of their minimum clock period. Clocking is thus very challenging, as SFQ circuits are more sensitive to setup failures that cause increases in clock-to-Q delay than their CMOS counterparts. It is therefore important to model and analyze clock-to-Q delay, particularly in the context of crossing clock domains where setup violations are expected. 

\section{Modeling Metastability}\label{sec:Metastability_modeling}


Let us first introduce a few relevant notations illustrated in Fig. \ref{fig:dff}(c). We denote the clock-to-Q delay of a DFF as $T_c$ and the time before the clock that the data pulse arrives at a DFF as $T_d$. We define $t_0$ as the time of the arrival of the data before the clock when the clock-to-Q delay starts to increase from its' nominal value. We also define $t_m$, where $t_m \leq t_0$, as the minimum data arrival time before the clock for which the corresponding output pulse is generated in the same clock cycle. Note that the DFF enters a metastable state once the clock pulse arrives if $T_d=t_m$. In this case, the clock-to-Q delay approaches infinity, i.e., the comparator does not know whether to trigger $J_2$ or $J_3$. However, similar to CMOS, any small perturbation will take the DFF out of this state. As a result, it will either trigger $J_3$ with delaying the output pulse until the arrival of the next clock pulse or $J_2$ with a high clock-to-Q delay. However, unlike in CMOS, the delayed output pulse will be generated as soon as the next clock pulse arrives, limiting the overall impact of metastability. Thus, unlike CMOS, the increased clock-to-Q delay is not the length of time the DFF remains in metastability, but instead caused by the superconducting nature of the circuit near the metastable state, i.e., as $T_d$ approaches $t_m$. This is described in more detail below.

\subsection{Reason for Increased Clock-to-Q Delay}

To gain a physical understanding of the relationship between $T_d$ and $T_c$, we further analyze the SFQ DFF. The output pulse of the DFF is delayed when the junction $J_{1}$ does not get enough time to undergo a $2\pi$ phase leap and thereby cannot flip the state of the $J_{1}-L_{2}-J_{2}$ loop before the arrival of the clock pulse. The time taken by $J_{1}$ to flip the state is derived below.

Consider the RCSJ model of $J_{1}$, as shown in Fig. \ref{fig:dff}(d) with a capacitor (C) and resistor (R) in parallel, the latter acting as a shunt to overdamp $J_1$. The total current which is the sum from Kirchoffs' laws is given by 

\begin{equation}\label{eq-1}
I(t) = I_{c} \sin \phi(t) + \frac{V(t)}{R} + C\diff{V(t)}{t}
\end{equation}

\noindent
Here, $V(t)$ is the voltage across $J_{1}$ and $I(t)$ the total current. $I_{c}$ and $\phi$ are the critical current and phase of $J_{1}$ respectively. Using flux to voltage conversion in Eq. \ref{eq-2}, we obtain the full current equation in Eq. \ref{eq-3}.
\begin{equation}\label{eq-2}
V(t)=\frac{\hbar}{2e} \diff{\phi(t)}{t}
\end{equation}
\begin{equation}\label{eq-3}
I(t) = I_{c} \sin \phi(t) + \frac{\hbar}{2eR} \diff{\phi(t)}{t} + C\frac{\hbar}{2e}\diff[2]{\phi(t)}{t}
\end{equation}
which is a second order nonlinear ordinary differential equation. Note that $e$ is the elementary charge and $\hbar$ is the reduced Planck's constant. $J_{1}$ is initially biased in superconducting state with $I(t)=I_{0}$, where $I_{c} \geq I_0$. Note that $I_{0}$ is primarily provided by the bias current source $I_{b}$. 

The phase $\phi(t)$ in this condition does not change with time.

\begin{equation}\label{eq-4}
\phi(t)= \phi_0 = \arcsin(\frac{I_0}{I_{c}})
\end{equation} 

\noindent
and hence, $V(t)$ is zero. With the arrival of the input pulse, $I(t)$ becomes $I_{1}$ with $I_{c} < I_{1}$, the phase grows with time and we can observe a nonvanishing voltage. The time required to increase this phase by an angle of $2\pi$ can result in one quantum flux being stored in the inductance loop. This time is denoted $t_0$. Since we use overdamped JJs in SFQ logic, we can ignore the RC time constant since it is much smaller than the intrinsic time constant of $J_{1}$. With this assumption, we rewrite Eq. \ref{eq-3} 
ignoring the second order term, as follows, 

\begin{equation}\label{eq-5}
\frac{\hbar}{2eR}\frac{d\phi(t)}{I_{1}-I_{c}\sin\phi(t)}=dt
\end{equation}
and integrate over a $2\pi$ change in $\phi(t)$ to obtain
\begin{equation}\label{eq-6}
t_0=\frac{\hbar}{2eR}\frac{2\pi}{\sqrt{I_{1}^{2}-{I_{c}}^{2}}}
\end{equation}

We can also integrate over an arbitrary time period $(0,T_d)$ with $T_d \leq t_0$ and obtain

\begin{equation}\label{eq-7}
\phi(T_d)=
\begin{cases}
       2\arctan\Bigg(a\cdot\tan{\frac{T_d\cdot b}{2\tau}}\Bigg)+ \phi_{0},  & \text{if } T_d\leq \frac{\pi\tau}{b}\\
    2\arctan\Bigg(a\cdot\tan{\frac{T_d\cdot b}{2\tau}}\Bigg)+ \phi_{0}+ 2\pi,              & \text{otherwise}
\end{cases}
\end{equation}
where $\tau=\frac{\hbar}{2eI_{c}R}$, $a=\sqrt{1-\left({\frac{I_{c}}{I_{1}}}\right)^{2}}$ and $b=\sqrt{{\left(\frac{I_{1}}{I_{c}}\right)^{2}}-1}$.
Note that $\phi(T_d)$ is an increasing function in $T_d$ with $\phi(0)=\phi_{0}$ and $\phi(t_0)=\phi_{0}+2\pi$. 

The output clock-to-Q delay is nominal when the input pulse arrives at time $t \geq t_0$. However, any pulse on the data input, if not given time $t_0$ before the clock, will result in some phase change less than $\phi(t_0)$ across $J_{1}$. Hence, when the clock pulse arrives, the resulting current $\delta I$ in $L_{2}$ will not result in a quantum of fluxon. As a result, the output pulse either comes out with a delay higher than the nominal value, as detailed in the next subsection, or does not come out until the arrival of the next clock pulse. 

To be more precise, the output comes out in the next clock cycle, if the pulse on the data input arrives later than $t_m$. When the input is later than $t_m$, the resulting current $\delta I$ produced in $L_{2}$ and passed into $J_{2}$ (including the shunt resistor and capacitor shown in Fig. \ref{fig:dff}(d)) becomes less than $J_2$'s critical current when the clock pulse arrives. Hence, $J_{3}$ will leap and no associated output pulse will be generated. 
However, since the inductance loop stores the flux, once the next clock pulse comes, $J_{2}$ will leap resulting in an output pulse. Hence, any late input pulse, specifically after $t_m$, delays the latency of the output pulse by one clock cycle.

\subsection{Modeling Increased Clock-to-Q Delay}\label{subsec:clk-to-q_delay}

Now, let us derive the temporal dynamics of the response time of the DFF, particularly when the input pulse arrives between $t_0$ and $t_m$. Without any input pulse, $J_{2}$ is biased at $I_{c}\sin{\theta_0}$ (similar to $J_{1}$) where $\theta_0$ is the static (superconducting) phase of $J_{2}$. With the $2\pi$ phase leap of $J_{1}$ and the arrival of the clock pulse, we inject additional current $\delta{I_d}$ and $\delta{I_{clk}}$ respectively into $J_{2}$, such that $I_{c}\sin{\theta_0}+\delta{I_d}+\delta{I_{clk}}$ becomes larger than $I_c$.

Note that $\delta{I}$ is upper bounded by ${I}_{max}$ which results from a complete $2\pi$ phase change in $J_1$. Moreover, we can substitute $I=I_{c}\sin{\theta_0}+I_{max}+\delta{I_{clk}}$ into Eq. \ref{eq-6} to get the nominal clock-to-Q delay of the flop. However, $\delta{I_d}$ starts to drop below $I_{max}$ when $T_d$ decreases below $t_0$, i.e., violates the setup time of the DFF. This decrease continues until $T_d$ becomes $t_m$ where $I_{c}\sin{\theta_0}+\delta{I_d}+\delta{I_{clk}}=I_{c}$ and $T_c$ approaches infinity. 

The phase change across $J_{1}$ computed as $\phi(T_d)$ is similar to the angular magnetic flux which is proportional to the current across the inductor $L_{2}$. Therefore, $\delta{I_d}$ can be written as $K_{1}{\phi(T_d)}$ with $K_1$ being an arbitrary constant. Referring to 
Eq. \ref{eq-6}
and linking the change in input current (phase) as a function of time of arrival of the input pulse $T_d$, the clock-to-Q delay can be modelled as 
\begin{equation}\label{eq-8}
T_c=f(T_d)=\frac{\hbar}{2eR}\frac{2\pi}{\sqrt{\left(I_x+K_1\phi(T_d)\right)^{2}-{I_c}^{2}}} + K_2
\end{equation}
\noindent
where $I_x=I_c\sin{\theta_0}+\delta{I_{clk}}$. $K_{2}$ has been introduced to model any output buffer delay and $\phi(T_d)$ is defined by Eq. \ref{eq-7}.

\subsection{Modeling MTBF}

Given the relationship between $T_d$ and $T_c$, we can now derive the equations for the failure rate of a one-flop DFF synchronizer in the presence of a data pulse whose arrival times are uncorrelated to the clock input. We denote the timing slack at the output of the synchronizer under consideration as $t_{r}$. This is the maximum time after the clock pulse that the synchronizer output is allowed to generate an output pulse. In our experiments, we set $t_{r}$ to be roughly 10\% higher the nominal clock-to-Q delay of the DFF, as is typical in standard cell libraries \cite{nkatam_timing}. 
Interestingly, if the predicted clock-to-Q delay exceeds the clock period $t_{clk}$, the output pulse appears earlier, directly after the next clock pulse, with negligible delay. This behavior does not cause any harm, because the handshaking protocol associated with synchronizers typically account for this potential increase in pulse latency \cite{gdatta_isec}.

\begin{figure}
\centering {
\includegraphics[width=9cm]{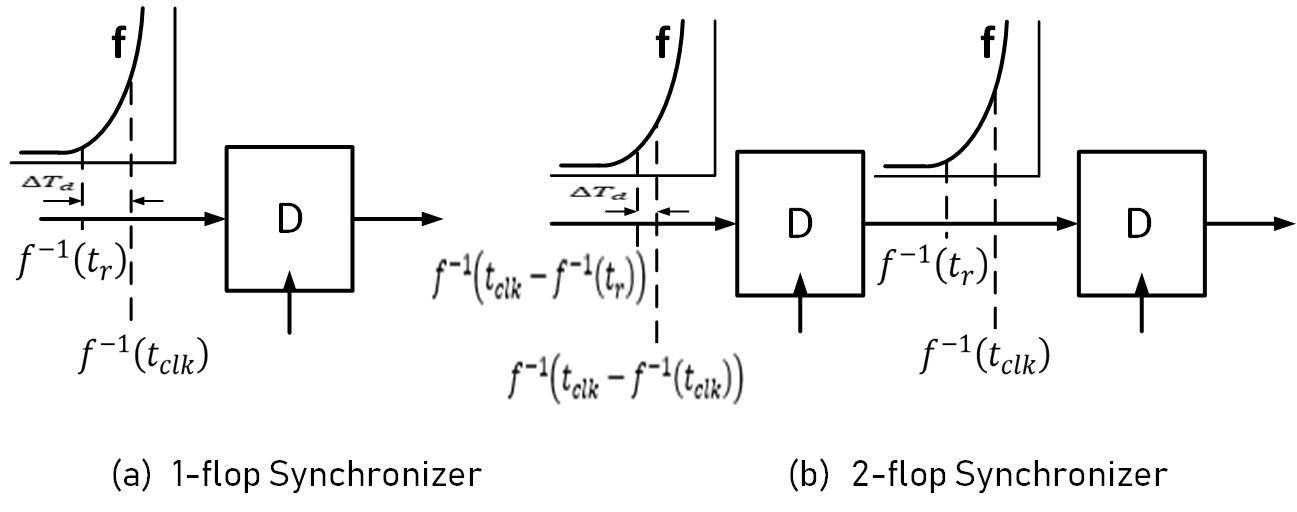}
\vspace{-0.7cm}
\caption{Transition window for failure $\Delta{T_d}$ of (a) one-flop and (b) two-flop synchronizer}
\label{fig:mtbf}
}
 \vspace{-.4cm}
\end{figure}

Thus the probability of failure can thus
be expressed as $p(\mathit{failure})=p(t_{r} \leq T_c \leq t_{clk})$. Using Eq. \ref{eq-8}, we obtain
\begin{equation}\label{eq-9}
p(\mathit{failure})=p(f^{-1}(t_{clk}) \leq T_d \leq f^{-1}(t_{r}))
\end{equation}
The probability of failure is, thus equal to the probability of a pulse arriving in the window $\Delta T_d = (f^{-1}(t_{clk}) - f^{-1}(t_{r}))$ illustrated in Fig \ref{fig:mtbf}(a). Assuming a clock frequency of $F_{c}$ and that the input data arrival time is uniformly distributed across the clock period, the probability of failure is
\begin{equation}\label{eq-10}
p(\mathit{failure})=F_{c}*(f^{-1}(t_{r})-f^{-1}(t_{clk}))
\end{equation}
Assuming that the DFF is operating at a frequency of $F_{d}$, the total number of failures per second will be $F_{c}*F_{d}*(f^{-1}(t_{r})-f^{-1}(t_{clk}))$.
The MTBF of a single flop synchronizer is simply the recipriocal of this value,
\begin{equation}\label{eq-11}
MTBF=\frac{1}{F_{c}*F_{d}*(f^{-1}(t_{r})-f^{-1}(t_{clk}))}
\end{equation}
The addition of a second FF to the synchronizer decreases the size of this window, as 
illustrated in Fig. \ref{fig:mtbf}(b), increasing the MTBF. Note that our modeling approach is similar to the analysis of multistage CMOS synchronizers in \cite{beer_MTBF} 
but our resulting equation is not closed-form. 
Finally we note that while MTBF grows rapidly as a function of the number of flip-flops the underlying function \textit{f} is not an exponential, in contrast to CMOS \cite{ginosar_sync, kinniment_tutorial}.

\section{Simulating Metastability}\label{sec:Metastability_simulate}

The MTBF is a function of several device parameters that can be extracted from JSIM simulations of a DFF. This section describes our simulation results and curve fitting to estimate these parameters and the resulting MTBF of various flop synchronizers.

\subsection{Simulation Setup and Results}

\begin{figure}[b]
\centering {
\includegraphics[width=9cm]{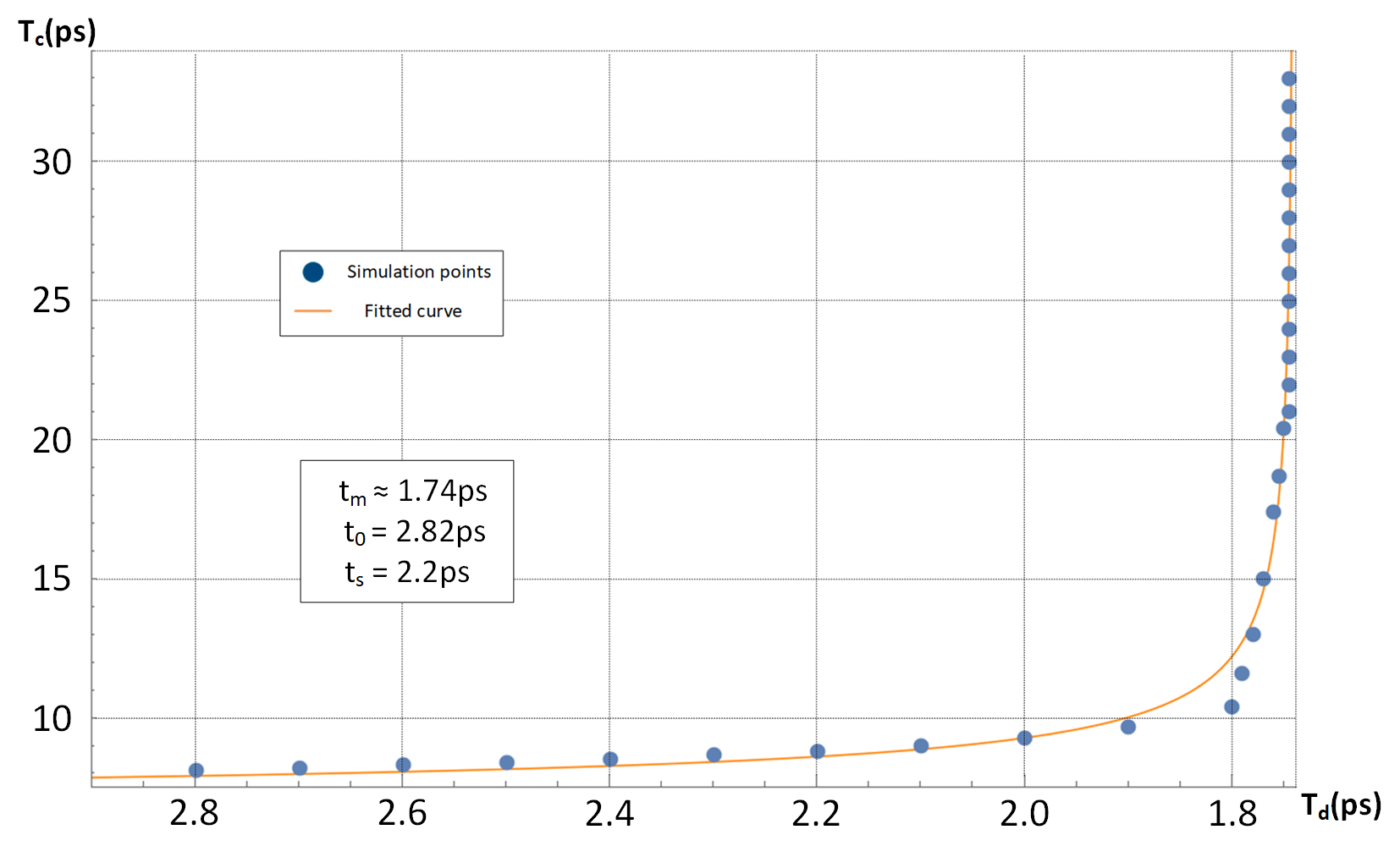}
\caption{Clock-to-Q delay ($T_c$) of a DFF as a function of the relative arrival time of the input pulse with respect to the clock ($T_d$). The dotted points indicate simulation results and the overlaid curve is the best fit of our proposed model.}
\label{fig:sim}
}
 \vspace{-.4cm}
\end{figure}

To observe the relationship between $T_d$ and $T_c$, we designed a custom DFF in the MIT LL $100\mu A/\mu m^{2}$ SFQ5ee process. We have kept $t_{clk}$ equal to the highest clock-to-Q delay that the simulator exhibits $({\sim}$39 $ps)$ to maximize the number of data points in the steep region, near $t_m$, we give to our curve fitting program.

The dotted points in Fig. \ref{fig:sim} show the simulation results of clock-to-Q delay $T_c$ of our designed DFF as a function of the relative timing of the data pulse $T_d$. The curve on the top of the dotted points is obtained by curve fitting to Eq. \ref{eq-8}. As we sweep $T_d$ towards the clock pulse, we observe that the clock-to-Q delay starts to increase from $T_d=t_0$ until $T_d$ reaches $t_m$ where the flop no longer captures the input data pulse. We have refined the precision of $t_m$ until $T_d$ increments reach the minimum time difference that the simulator can resolve. Simulated values of $t_0$, $t_m$ and the setup time of our DFF $(t_s)$ are shown in Fig. \ref{fig:sim}.

\begin{table}[!ht]
\caption{Failure windows and MTBFs of different synchronizers}
\begin{center}
\begin{tabular}{|c|c|c|c|}
\hline
\textbf{Number of} & \textbf{Clock freq.} & \textbf{Failure transition} & \multirow{2}{*}{\textbf{MTBF}}  \\
\textbf{flops} &  $F_{c} $\textbf{ (GHz)} & \textbf{window $\Delta{T_d}$ (ps)} & \\
\hline
 \multirow{3}{*}{$1$} & $25$ & $0.41$ & $0.039 \mu s$  \\
\cline{2-4}
& $30$ & $0.405$ & $0.027 \mu s$\\
\cline{2-4}
 & $35$ & $0.405$ & $0.02 \mu s$ \\
\hline
\multirow{3}{*}{$2$} & $25$ & $2.3\times10^{-20}$ & $8.05\times 10^{5} yrs$  \\
\cline{2-4}
& $30$ & $5.3\times10^{-12}$ & $1.8\times 10^{3} s$ \\
\cline{2-4}
 & $35$ & $1.4\times10^{-7}$  & $0.036 s$ \\
\hline
\end{tabular}
\label{mtbf_values}
\end{center}
\end{table}
\subsection{MTBF Computation}

To compute MTBF from Eq. \ref{eq-11}, we used typical values of $F_{c}=25$ $GHz$, $F_{d}=2.5$ $GHz$ and $t_{r}=8$ $ps$. This is because the clock frequency of the current state-of-the-art SFQ processor is around $25$ $GHz$\cite{core_1b} and we assume we have a data pulse once every ten clock cycles. We have kept $t_{r}$ at $8 ps$, because it is roughly $10\%$ higher the nominal clock-to-Q delay of our DFF. 
To evaluate the function $f(T_d)$ as defined in Eq. \ref{eq-8}, we set the values of $I_{c}$ and $R$ as used in our design and used curve fitting to estimate the other device parameters, namely $I_{1}$, $I_x$, $\phi_{0}$, $K_{1}$, and $K_{2}$. This was motivated because $I_1$, $I_x$, and $\phi_0$ are difficult to otherwise evaluate and $K_1$ and $K_2$ involve non-linear effects that are not captured by our model. We ensured that the fitted values of these parameters are realistic. Our fitted function $f(T_d)$ has a Root Mean Square Error (RMSE) of $0.32$ $ps$ which is around $1\%$ of the range of the dependent variable, $T_c$. 
RMSE is defined as
\begin{equation}\label{eq-12}
RMSE=\sqrt{\frac{\sum_{i=1}^{n} [t_{c_i}-f(t_{d_i})]^{2}}{n}}
\end{equation}
and $(t_{c_i},t_{d_i})$ $\forall i=\{1,2,...,n\}$ are the $n$ simulation points.

Plugging the assumed and fitted values described above in Eq. \ref{eq-11}, we obtain an MTBF of $0.027 \mu{s}$ for our one-flop synchronizer. Table \ref{mtbf_values} illustrates how we can improve this value by the addition of one more flop. Table \ref{mtbf_values} also describes the significant degradation in MTBF when the clock frequency $F_{c}$ is increased.\footnote{As mentioned earlier, the data frequency $(F_{d})$ is always kept at $10\%$ of the clock frequency.} For a clock frequency of 30 GHz, the MTBF of our two-flop synchronizer in the SFQ5ee process technology is estimated to be \textbf{$8.05\times10^{5}$} years. 

\section{Conclusions}\label{sec:conc}

In this paper, we have derived an analytical model for metastability in SFQ from first principles from which we derived an equation for MTBF of DFF-based SFQ synchronizers.  
We applied this model to the 
MIT LL $100 \mu A/\mu m^{2}$ SFQ5ee process by curve fitting our model to detailed JSIM simulations of a DFF designed in this process. The model fits our simulation results well showing low RMSE. Our model predicts that while a single DFF would lead to low MTBF, the standard back-to-back two-flop synchronizer, operating at $25$ $GHz$, has an estimated MTBF of ${\sim}10^{6}$ years. 

Our future work includes the design of high-throughput synchronizers and demonstrating their use in large-scale SFQ designs. It would also be interesting to analyze the rate of growth of MTBF as a function of the number of flip-flops in an SFQ synchronizer and compare it to the standard exponential growth observed in CMOS.

\section{Acknowledgement}

The research is based upon work supported by the Office of the Director of National Intelligence (ODNI), Intelligence Advanced Research Projects Activity (IARPA), via the U.S. Army Research Office grant W911NF-17-1-0120. The views and conclusions contained herein are those of the authors and should not be interpreted as necessarily representing the official policies or endorsements, either expressed or implied, of the ODNI, IARPA, or the U.S. Government. The U.S. Government is authorized to reproduce and distribute reprints for Governmental purposes notwithstanding any copyright notation herein.

\bibliographystyle{IEEEtran}
\bibliography{IEEEabrv,bibliography}

\end{document}